\documentclass[preprint,12pt]{elsarticle}

\usepackage{graphicx}
\usepackage{url}
\usepackage{lscape}
\usepackage{multirow}
\usepackage{algpseudocode}
\usepackage{algorithm}
\usepackage{amsthm}
\usepackage{amsmath}
\usepackage{amssymb}
\usepackage{float}
\usepackage{subfig}
\usepackage[nodots]{numcompress}

\journal{.}

\begin{document}

\begin{frontmatter}



\title{MILP and Max-Clique based heuristics for the Eternity II puzzle} %GVB problem}


\author[1]{Fabio Salassa}
%\ead{fabio.salassa@polito.it}
\author[2]{Wim Vancroonenburg}
%\ead{wim.vancroonenburg@kahosl.be}
\author[2]{Tony Wauters\corref{cor1}}
\ead{tony.wauters@cs.kuleuven.be}
\author[1]{\\Federico Della Croce}
%\ead{federico.dellacroce@polito.it}
\author[2]{Greet Vanden Berghe}

\cortext[cor1]{Corresponding author}

\address[1]{Politecnico di Torino, DIGEP, Corso Duca degli Abruzzi 24, 10129 Torino, Italy}
\address[2]{KU Leuven, Department of Computer Science, CODeS \& iMinds-ITEC,\\ Gebroeders De Smetstraat 1, 9000 Gent, Belgium}

\begin{abstract}
The present paper considers a hybrid local search approach to the Eternity II puzzle 
and to unsigned, rectangular, edge matching puzzles in general.
Both an original mixed-integer linear programming (MILP) formulation and a novel Max-Clique formulation are presented for this NP-hard problem.
Although the presented formulations remain computationally intractable for medium and large sized instances, they can serve as the basis for developing heuristic decompositions and very large scale neighbourhoods. As a side product of the Max-Clique formulation, new hard-to-solve instances are published for the academic research community.
Two reasonably well performing MILP-based constructive methods are presented and used for determining the initial solution of a multi-neighbourhood local search approach.
Experimental results confirm that this local search can further improve the results obtained by the constructive heuristics and is quite competitive with the state of the art procedures.

\end{abstract}

\begin{keyword}
Edge matching puzzle \sep hybrid approach \sep local search


\end{keyword}

\end{frontmatter}

% \linenumbers

\section{Introduction}
\label{sec:Intro}

\begin{figure}[h]
\centering
		\includegraphics[scale=0.25]{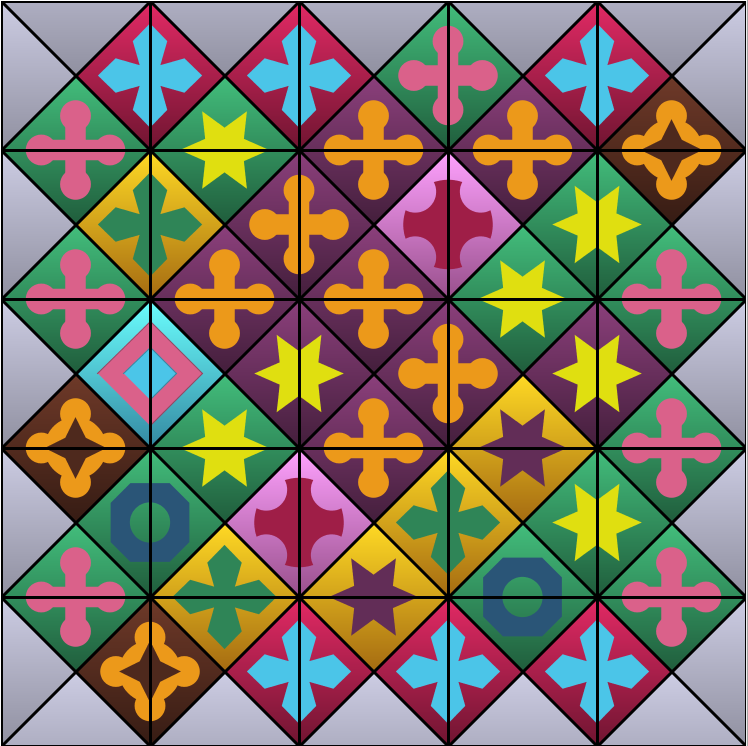}
	\caption{Solution to an Eternity II-like edge matching puzzle of size $5 \times 5$ (Image generated with the Eternity II editor, \textit{http://sourceforge.net/projects/eternityii/}, accessed on 24 January 2014)} %GVB November 2011)}
\label{fig:e2image}
\end{figure}

\noindent 
The Eternity II puzzle (EII) is a commercial edge matching puzzle in which 256 square tiles with four \emph{coloured} edges must be arranged on a $16\times 16$ grid such that all tile edges are matched.
%\\
%\\
%The Eternity II puzzle (EII) is an edge matching puzzle with 256 square tiles that need to be placed in a grid with 16 rows and 16 columns, each tile having four edges with a coloured pattern.
%A solution to this puzzle requires that each tile is placed in a unique position in the grid such that the edge shared between any two adjacent tiles is matched.
In addition, a complete solution requires that the `grey' patterns, which appear only on a subset of the tiles, should be matched to the outer edges of the grid.
%To illustrate this, an example solution to a smaller $5 \times 5$ version is provided in Figure \ref{fig:e2image}.
An illustration of a complete solution for a small size puzzle, $5 \times 5$, is provided in Figure \ref{fig:e2image}.

The EII  puzzle %GVB problem 
%is a commercial edge matching puzzle, 
was created by Christopher Monckton and released by the toy distributor Tomy UK Ltd.~in July 2007.
Along with the puzzle release, a large cash prize of 2 million USD was announced to be awarded to the first person who could solve the puzzle.
As can be expected, this competition attracted considerable attention.
Many efforts were made to tackle this challenging problem, yielding interesting approaches and results. 
However, no complete solution has ever been generated.
Meanwhile, the final scrutiny date for the cash price, 31 December 2010, has passed, leaving the large money prize unclaimed.
\\
\\
The EII puzzle belongs to the more general class of Edge Matching Puzzles, which have been shown to be NP-complete \cite{DD07}.
Many approaches to Edge Matching Puzzles are now available in the literature. 
Constraint programming approaches \cite{BB08,SD08} have been developed, in addition to metaheuristics \cite{CCCHSO10,SD08,WC10}, backtracking \cite{VE08} and evolutionary methods \cite{MGS09}. 
Other methods translate the problem into a SAT formulation and then solve it with SAT solvers \cite{ABFM08,HE08}.
An extensive literature overview on the topic is provided in \cite{WVV12}, while \cite{KPS08} provides a survey on the complexity of other puzzles.
\\
\\
The present paper introduces a novel Mixed-Integer Linear Programming (MILP) model and a novel Max-Clique based formulation for EII-like puzzles of size $n \times n$. %GVB problem. 
Both formulations serve as components of heuristic decompositions used in 
%two distinct constructive methods and improved by 
a multi-neighbourhood local search approach.
The remainder of the paper is structured as follows.
Section \ref{sec:models} presents the MILP and Max-Clique formulations. 
In Section \ref{sec:e2solutionapproach}, several hybrid heuristic approaches are introduced. Computational results are presented in Section \ref{sec:e2results}. Final conclusions are drawn in Section \ref{sec:e2conclusion}.

\section{Problem formulations}
\label{sec:models}

\subsection{Mixed integer linear programming formulation}
\label{subsec:MILP}
\noindent
A novel mixed-integer linear programming model was developed for the EII puzzle problem.
The following notation will be used. The puzzle consists of an $n\times n$ square onto which $n^2$ tiles need to be placed.
The index $t = 1 \ldots n^2$ is used to refer to tiles.
The indices $r = 1 \ldots n$, $c = 1 \ldots n$ denote the rows, resp. columns, of the puzzle board.
The index $\alpha = 0\ldots 3$ refers to the rotation of the tile, i.e.\ $\alpha = 0$ means not rotated,\ $\alpha = 1$ means rotated \emph{clockwise} over $90^\circ$, etc.\ 
The coefficient $CT_{t,\alpha,l}$ (resp. $CB$, $CL$, $CR$) is equal to $1$ if tile $t$ has colour $l$ at its top (resp. bottom, left, right) position when rotated by $\alpha$.\\

%The decision variables $x_{t,r,c,\alpha}$ are defined as follows:
\noindent
The decision variables of the MILP model are defined as follows:
{\allowdisplaybreaks
\begin{align*}
x_{t,r,c,\alpha} = & \begin{cases} 
1 & \textrm{if tile $t$ is placed in row $r$, column $c$ with rotation $\alpha$},\\
0 & \textrm{otherwise}.
\end{cases}\\
%\end{align}
%The decision variables $h_{r,c}$ and $v_{r,c}$ are defined as follows:
h_{r,c} = & \begin{cases} 1 & \textrm{if the right edge of position $(r,c)$ is \emph{unmatched}},\\
0 & \textrm{otherwise}.
\end{cases}\\
v_{r,c} = & \begin{cases} 1 & \textrm{if the bottom edge of position $(r,c)$ is \emph{unmatched}},\\
0 & \textrm{otherwise}.
\end{cases}
\end{align*}

\noindent
The model is then defined as follows:
\begin{align}
& \textrm{Min}\ \sum_{r=1}^{n} \sum_{c=1}^{n-1} h_{r,c} + \sum_{r=1}^{n-1} \sum_{c=1}^{n} v_{r,c}\label{eq:m2obj}\\
s.t. & \nonumber\\
&\sum_{r=1}^n \sum_{c=1}^n \sum_{\alpha=0}^3 x_{t,r,c,\alpha}    = 1  \quad \forall t = 1, \ldots, n^2\label{eq:m1eachTileAssigned}\\
&\sum_{t=1}^{n^2}\sum_{\alpha=0}^3 x_{t,r,c,\alpha} = 1\quad\forall r=1,\ldots, n,\;c=1,\ldots, n \label{eq:m1oneTilePerPos}\\
&\sum_{t=1}^{n^2}{\sum_{\alpha=0}^3{CR_{t,\alpha,l} x_{t,r,c,\alpha}} } - \sum_{t=1}^{n^2}{\sum_{\alpha=0}^3{CL_{t,\alpha,l} x_{t,r,c+1,\alpha}} } \leq h_{r,c}\nonumber\\
&\quad\forall r=1,\ldots,n,\;c=1,\ldots ,n-1,\;l=1,\ldots ,L\label{eq:m2horizontalEdge1}\\
-&\sum_{t=1}^{n^2}{\sum_{\alpha=0}^3{CR_{t,\alpha,l} x_{t,r,c,\alpha}} } + \sum_{t=1}^{n^2}{\sum_{\alpha=0}^3{CL_{t,\alpha,l} x_{t,r,c+1,\alpha}} } \leq h_{r,c}\nonumber\\
&\quad\forall r=1,\ldots,n,\; c=1,\ldots ,n-1,\;l=1,\ldots ,L\label{eq:m2horizontalEdge2}\\
&\sum_{t=1}^{n^2}{\sum_{\alpha=0}^3{CB_{t,\alpha,l} x_{t,r,c,\alpha}} } - \sum_{t=1}^{n^2}{\sum_{\alpha=0}^3{CT_{t,\alpha,l} x_{t,r+1,c,\alpha}} } \leq v_{r,c}\nonumber\\
&\quad\forall r=1,\ldots,n-1,\;c=1,\ldots ,n,\;l=1,\ldots , L\label{eq:m2verticalEdge1}\\
-&\sum_{t=1}^{n^2}{\sum_{\alpha=0}^3{CB_{t,\alpha,l} x_{t,r,c,\alpha}} } + \sum_{t=1}^{n^2}{\sum_{\alpha=0}^3{CT_{t,\alpha,l} x_{t,r+1,c,\alpha}} } \leq v_{r,c}\nonumber\\
&\quad\forall r=1,\ldots,n-1,\;c=1,\ldots ,n,\;l=1,\ldots ,L\label{eq:m2verticalEdge2}\\
&\sum_{t=1}^{n^2}{\sum_{\alpha=0}^3{CT_{t,\alpha,0}\cdot x_{t,0,c,\alpha}} } = 1\quad\forall c=1,\ldots, n\label{eq:m1topGray}\\
&\sum_{t=1}^{n^2}{\sum_{\alpha=0}^3{CB_{t,\alpha,0}\cdot x_{t,n-1,c,\alpha}} } = 1\quad\forall c=1,\ldots, n\label{eq:m1bottomGray}\\
&\sum_{t=1}^{n^2}{\sum_{\alpha=0}^3{CL_{t,\alpha,0}\cdot x_{t,r,0,\alpha}} } = 1\quad\forall r=1,\ldots, n\label{eq:m1leftGray}\\
&\sum_{t=1}^{n^2}{\sum_{\alpha=0}^3{CR_{t,\alpha,0}\cdot x_{t,r,n-1,\alpha}} } = 1\quad\forall r=1,\ldots, n\label{eq:m1rightGray}\\
&x_{t,r,c,\alpha} \in \{0,1\}\nonumber\\
&\forall t= 1,\ldots, n^2, \;r=1,\ldots, n, \;c=1,\ldots, n,\;\alpha = 0,\ldots, 3\label{eq:m2xvars}\\
&0\leq h_{r,c} \leq 1\quad \forall r=1,\ldots, n, \;c=1,\ldots,n-1\label{eq:varh}\\
&0\leq v_{r,c} \leq 1\quad \forall r=1,\ldots, n-1, \;c=1,\ldots,n\label{eq:varv}
 \end{align}

\noindent
The objective function (Expression \ref{eq:m2obj}) minimises the number of unmatched edges in the \textit{inner} region of the puzzle.
Constraints (\ref{eq:m1eachTileAssigned}) indicate that each tile must be assigned to exactly one position, with one rotation.
Constraints (\ref{eq:m1oneTilePerPos}) require that exactly one tile must be assigned to a position.
The edge constraints (\ref{eq:m2horizontalEdge1}) and (\ref{eq:m2horizontalEdge2}) force the $h_{r,c}$ variables to take on the value 1 if the tiles on positions $(r,c)$ and $(r,c+1)$ are unmatched.
Similarly, constraints (\ref{eq:m2verticalEdge1}) and (\ref{eq:m2verticalEdge2}) do the same for the vertical edge variables $v_{r,c}$.
Finally, constraints (\ref{eq:m1topGray}) - (\ref{eq:m1rightGray}) ensure that the border edges are matched to the gray frame (colour $l=0$).
\\
\\
We point out that constraining the objective function to zero (i.e.\ no unmatched edges allowed), turns the model into a feasibility problem where every feasible solution is also optimal.
However, preliminary testing showed that the latter model is only relevant for very small size problem instances.
If the MILP solver needs to be stopped prematurely on the feasibility model, no solution is returned.

\subsection{Clique formulation}
\label{subsec:clique}

\noindent
The EII puzzle itself is a decision problem and can be modelled as (i.e.\ it reduces to) the well known decision version of the \textit{clique} problem \cite{BBPP99} as follows.
Given a parameter $k$ and an undirected graph $G=(V,E)$, the \textit{clique} problem calls for finding a subset of pairwise adjacent nodes, called a clique, with a cardinality greater than or equal to $k$.
Let the nodes of the graph correspond to variables $x_{t,r,c,\alpha}$ from the formulation introduced in Section \ref{subsec:MILP}. Each node thus represents a tile in a given position on the puzzle and with a given rotation $\alpha$. 
The nodes are connected $\mathrm{iff}$ there is no conflict between the nodes in the puzzle.
Possible causes of conflicts are:
\begin{itemize}
	\item unmatching colors for adjacent positions
	\item same tile assigned to different positions
	\item same tile assigned to the same position with different rotations
	\item different tiles assigned to the same position.
\end{itemize}
The objective is to find a clique of size $n^2$, where $n$ is the size of the puzzle. %GVB (for squared instances).
%Table \ref{tab:ProblemInstancesAndTheirPropertiesClique} reports the number of nodes and the approximative number of edges for various sizes of the puzzle.
%For all instances the average graph density is 90\%.
\subsection{Comparison of the MILP model and the Clique model}
The applicability of the MILP model and the Clique model is investigated in what follows. Initial testing was performed on a set of small puzzle instances, ranging from $3\times3$ up to $8\times8$ (refer to Section \ref{sec:e2results} for more information on these instances).
The MILP model was implemented using CPLEX $12.6$.
A state of the art heuristic was used for solving the maximum clique problem \cite{GLP08}, 
kindly provided by its authors.
The heuristic has only one parameter, i.e.\ the number of selections $q$.
The computing time of the algorithm is linear with respect to $q$. 
We tested the heuristic with $q \in \left\{ 1,000,000;\ 10,000,000;\ 50,000,000  \right\}$.
Both the MILP model and the Max Clique heuristic were tested  on a modern desktop pc\footnote{Intel Core i7-2600 CPU@3.4Ghz}.

Table \ref{tab:cliqueResults} shows the results obtained with the max clique formulation and the MILP formulation.
\begin{table}
\scriptsize
	\centering
		\begin{tabular}{|l|c|c|c|c|c|c|}
		\hline
Puzzle size			 & 3x3    & 4x4   & 5x5   & 6x6   & 7x7  & 8x8  \\
			 \hline 
Optimal solution & 12 & 24 & 40 & 60 & 84 & 112 \\
\hline \hline
			\textbf{Clique} & & & & & & \\
			\hline
Number of nodes & 30 & 138 & 478 & 1290  & 2910  & 5770\\
\hline
Number of edges & 291 & 6707 & 86184 & 674512 & 3620565 & 14751304 \\
\hline
Graph density & 66.9$\%$ & 70.9$\%$ & 75.6$\%$ & 81.1$\%$ & 85.5$\%$ & 88.6$\%$ \\
\hline
 & $E$ ($T$) & $E$ ($T$) & $E$ ($T$) & $E$ ($T$) & $E$ ($T$) & $E$ ($T$)   \\
$q=1,000,000$ & \textbf{12} (0.2) & \textbf{24} (0.3) & \textbf{40} (0.8) & 57 (1.4) & 78 (2.5) & 102 (4.8) \\
$q=10,000,000$ & \textbf{12} (1.9) & \textbf{24} (3.8) & \textbf{40} (7.6) & \textbf{60} (13.5) & 80 (22.4) & 106 (35.7) \\
$q=50,000,000$ & \textbf{12} (9.3) & \textbf{24} (19.3) & \textbf{40} (37.8) & \textbf{60} (67.0) & 82 (110.5) & 106 (172.4) \\
		\hline \hline
			\textbf{MILP} & & & & & & \\
		\hline
			Number of variables & 336 & 1048 & 2540 & 5244  & 9688  & 16496\\
		\hline
			Number of constraints & 126 & 288 & 630 & 1056  & 1638  & 2624\\
		\hline
		 & $E$ ($T$) & $E$ ($T$) & $E$ ($T$) & $E$ ($T$) & $E$ ($T$) & $E$ ($T$)   \\
1 thread, 3600s time  & \textbf{12} (0.1) & \textbf{24} (0.1) & \textbf{40} (16.7) & \textbf{60} (243.5) & 81 (3600.0) & 103 (3600.0) \\
4 threads, 3600s time  & \textbf{12} (0.1) & \textbf{24} (0.1) & \textbf{40} (1.2) & \textbf{60} (158.5) & 81 (3600.0) & 103 (3600.0) \\
\hline
		\end{tabular}
	\caption{Results obtained with the maximum clique formulation solved with the algorithm by \cite{GLP08}, and the MILP formulation solved with CPLEX, on small size edge matching puzzle instances.}
	\label{tab:cliqueResults}
\end{table}
For each instance, we report for the clique formulation: the number of nodes, the number of edges, the optimal solution (namely the max.~number of matching edges), the density, the best number of matching edges $E$ and the average computing time $T$ (in seconds) for $10$ runs.
The number of variables and constraints is reported for the MILP formulation. 
The solutions depicted in bold are optimal.
The results show that instances up to size $6\times6$ can be easily solved using a state of the art maximum clique algorithm. 
Instances of size $7\times7$ could not be solved completely, even when the algorithm was executed with higher values of $q$ and more runs.
The MILP is also able to solve up to size $6\times6$. However, the clique formulation is significantly faster from size $6\times 6$ upwards.

Note that the edge matching puzzles correspond with large, difficult clique instances, for which current state-of-the-art max clique solvers are not able to find the optimal solution.
We provide the corresponding max clique instances of the $3\times3$ to $9\times9$ instances to the academic community\footnote{The instances can be downloaded from \url{https://dl.dropbox.com/u/24916303/Graph_Pieces.zip}. A generator for these instances is available upon request from the authors.}. Larger size instances are hard to manage. The $10\times10$ graph file, for example, is larger than 1 GB.

\section{Solution Approaches}
\label{sec:e2solutionapproach}

\noindent
Both the MILP model and the clique formulation presented in the previous section proved to be computationally intractable for medium sized instances. The size appears to be restricted to $7\times7$ and $8\times8$ when the execution time is limited to one day. The true EII puzzle instance is still far beyond the grasp of these models.
However, these models can serve as the basis for some well performing heuristics, presented in the following paragraphs.

\subsection{MILP-based greedy heuristic}
\label{subsec:MilpGreedy}

\noindent
A MILP-based greedy constructive heuristic has been developed for the problem studied here.
The heuristic is based on subproblem optimisation.
The puzzle is divided in regions, e.g.\ by considering individual rows/columns or rectangular regions.
Regions are then consecutively constructed by employing a variant of the MILP model presented in Section \ref{sec:models}.

First we introduce the notion of a partial solution $S^\ast = T^\ast \mapsto R^\ast$, in which a subset $T^\ast$ of the tiles $T = \{ 1,\ldots, n^2\}$ have been assigned to a subset $R^\ast$ of the positions $R = \{(r,c) | r = 1,\ldots,n, c = 1,\ldots, n\}$.
%Using this notation, $S^\ast(t)$ denotes the assignment of tile $t$ ($t\in T^\ast$) to a position $(r,c) \in R^\ast$.\\
%\noindent
Given a partial solution $S^\ast$, model (\ref{eq:m2obj})--(\ref{eq:varv}) can be modified such that it only considers the positions in a region $R'\subseteq R\backslash R^\ast$, and it only aims to assign tiles $T' \subseteq T\backslash T^\ast$ (i.e.\ tiles that have not been assigned elsewhere).
In addition, we restrict $R'$ to a rectangular region, denoted by $(r'_{min},c'_{min})\times (r'_{max},c'_{max})$ (i.e.\ the min/max positions of the region).\\
Figure \ref{fig:partialmip} illustrates how model (\ref{eq:m2obj})--(\ref{eq:varv}) can be modified to solve a region $(1,4)\times(4,8)$, given a partial solution on $(1,1)\times(4,4)$.
In this example, it is required to select $16$ of the available $64-16 = 48$ tiles (16 tiles are already assigned to region $(1,1)\times(4,4)$) in such a way that
the unmatched edges are minimised.
Hence, in order to consider region $(1,4) \times (4,8)$, the objective function is modified as follows
\begin{align}
& \textrm{Min}\ \sum_{r=1}^{4} \sum_{c=4}^{7} h_{r,c} + \sum_{r=1}^{3} \sum_{c=4}^{8} v_{r,c}
\end{align}
Only $16$ of the remaining $48$ tiles must be selected and assigned to the region and therefore constraints 
(\ref{eq:m1eachTileAssigned}) are modified as follows.
Note that the inequality indicates that not all tiles will be selected.
\begin{align}
&\sum_{r=1}^4 \sum_{c=4}^8 \sum_{\alpha=0}^3 x_{t,r,c,\alpha}    \leq 1  \quad \forall t \in T\backslash T^\ast
\end{align}
Similarly, constraints (\ref{eq:m1oneTilePerPos}--\ref{eq:varv}) 
are also suitably modified in order to take into account the specific region to be considered.
Note that the edge constraints forcing the values of the $h_{r,c}$ and $v_{r,c}$ variables also hold for rows and columns matching the boundaries of previously solved regions.
This enables building a solution with only a few unmatched edges between region boundaries.

\begin{figure}
\centering
\includegraphics[width=0.65\textwidth]{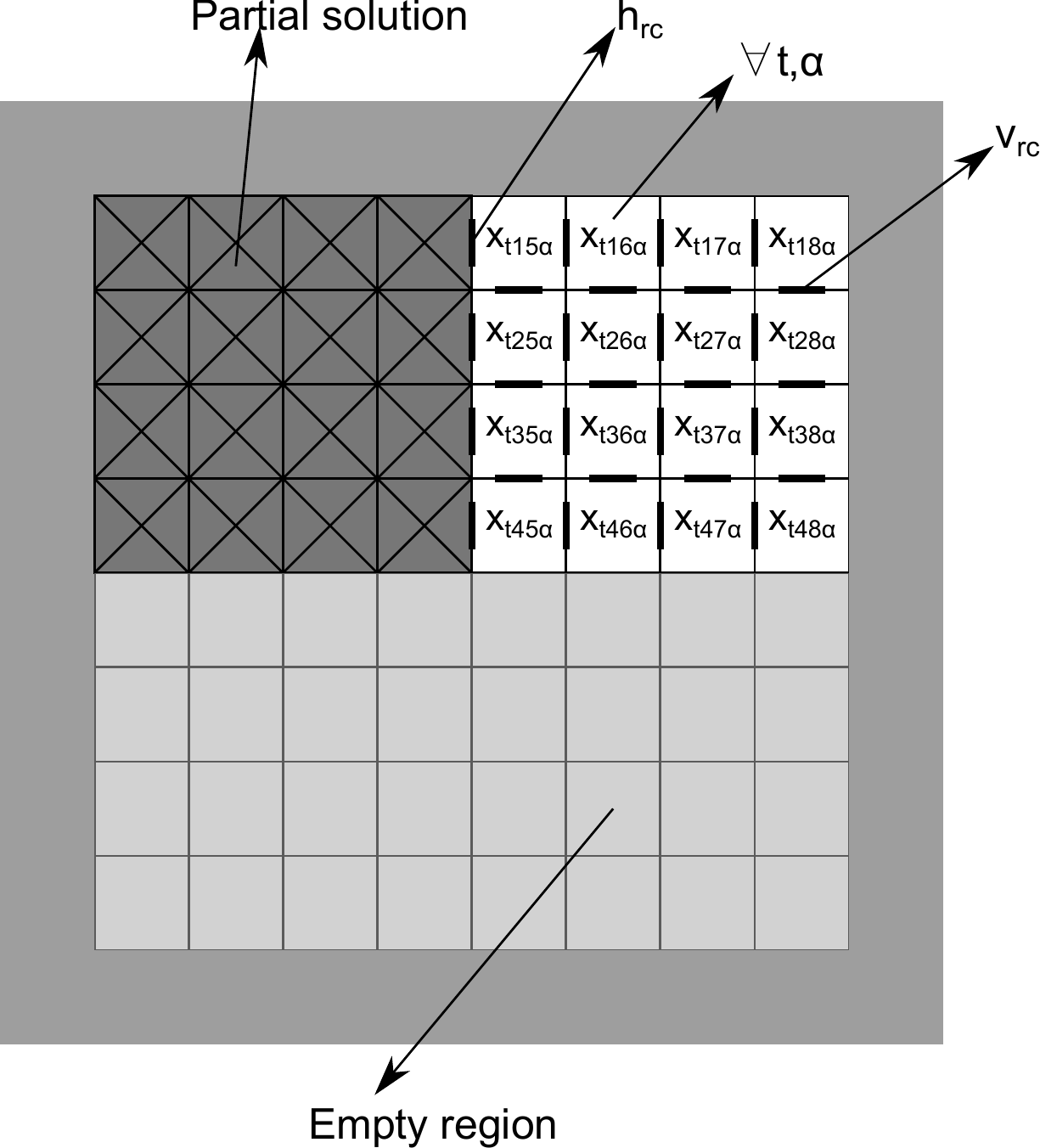}
\caption{Illustration of how model (\ref{eq:m2obj})--(\ref{eq:varv}) can be modified such that it only considers the positions in a region $(1,4)\times(4,8)$, given a partial solution $S^\ast$ on region $(1,1)\times(4,4)$.\label{fig:partialmip}
}
\end{figure}
This partial optimization model can be applied to solve all regions sequentially, thus constructing the final complete solution.
%Subproblems are generated by considering model (\ref{eq:m2obj})--(\ref{eq:varv}). 
%modified in such a way that adjacent regions of 
%fixed size and shape are taken into account.  %in the puzzle to be assigned.
%The proposed procedure considers subsets of the decision variables, as in classical greedy algorithms.  These subsets correspond to the variables in the considered regions, which are optimised 
%in a constructive way.
Initially $R_1$ is optimised, after which region $R_2$, disjoint from region $R_1$, is optimised, and so on.
The variables corresponding with the region are then optimally assigned by the MILP solver.
Algorithm \ref{alg:greedy} presents pseudocode of this approach. 
\begin{algorithm}
\small
\caption{Greedy heuristic}
\begin{algorithmic}
\Require $\mathcal{R} = \{R_1,R_2,R_3, \ldots R_K\}$ \Comment{A decomposition of $R$ in $K$ regions $R_i$}
\State $T^\ast_0 \gets \emptyset, R^\ast_0 \gets \emptyset, S^\ast_0 \gets \emptyset$ \Comment{The initial partial solution $S^\ast$ has no tiles assigned}
\For{$i = 1,\ldots,K$}
\State $S^\ast_i, T^\ast_i, R^\ast_i \gets $ Apply MILP model to region $R_i$, given $S^\ast_{i-1},T^\ast_{i-1}$ and $R^\ast_{i-1}$.
\State\Comment{Get a new partial solution $S^\ast_i$, with tiles assigned to $R_i$} 
\EndFor
\State \Return $S^\ast_K$
\end{algorithmic}
\label{alg:greedy}
\end{algorithm}

For each puzzle's size, differently sized subsets of tiles have been tested to assess the quality of the approach.
Preliminary tests of regions varying from 2 by 2 tiles (size 4) to 16 by 2 tiles (size 32) have been performed on the EII puzzle instance. 
This preliminary analysis revealed that the CPU time required at each iteration of the greedy heuristic limits the subset size to 32 tiles. This roughly corresponds to 32500 MILP variables for the first region of the real EII puzzle.
%Overall the best performing configuration of region size was solving 1 to 2 rows at each %iteration, for each puzzle instance.
Clearly, an increased number of tiles leads to better results. However, more CPU time is needed to compute the optimal solution, limiting the use in any hybrid framework.
%Indeed, as we set a limited time for each puzzle instance, the best trade-off seems to limit the %size of regions to 32 tiles. 
%Finally, in order to generate different solutions, an inner tile is randomly assigned to an inner position before the greedy heuristic starts.

\subsection{MILP-based backtracking constructive heuristic}
\label{subsec:MilpBacktrack}

\noindent
A backtracking version of the greedy heuristic has also been developed.
The main idea, namely building a complete solution by constructing optimal regions, is the same as for the greedy heuristic.
The backtracking version, however, restricts the optimal value of each subproblem to zero. All tiles in the region should match both internally and with respect to the tiles outside the region.
Whenever a subproblem is determined to be infeasible (i.e.\ no completely edge matching region can be constructed), the procedure backtracks to the previous region in order to find
a new assignment in that region.
This may afterwards enable constructing a feasible assignment in the next region.
If not, then the process is repeated until the backtracks are sufficient to find a complete solution. 

Model (\ref{eq:m2obj})--(\ref{eq:varv}), suitably modified, is again used to build partial solutions.
Let $R_i$ be the current region considered by the procedure and $S_{i}^\ast$ the related partial solution once the corresponding MILP model is solved.
Whenever the lower bound of the MILP model related to region $R_i$ is detected to be greater than zero, optimisation of region $R_k$ is stopped.
Instead, the previous region $R_{i-1}$ is reconsidered in order to obtain a new partial solution 
$S_{i-1}'^\ast$, again with value $0$. Let $X_{i-1}^*$ be the set of variables ${x}_{t,r,c,\alpha}$
having value $1$ in solution $S_{i-1}^*$.
%Let ${x^*}_{t,r,c,\alpha}$ be the value of variable $x_{t,r,c,\alpha}$ in solution $S_k^*$.
%In the search for solution $S_{k-1}'$,
The previous partial solution $S_{i-1}^\ast$ must be cut off when searching for solution $S_{i-1}'^\ast$. The following new constraint is added to the model:
\begin{equation}
\sum_{\hspace*{2cm} t,r,c,\alpha \; : \; x_{t,r,c,\alpha} \in X_{i-1}^*}  x_{t,r,c,\alpha} \leq |X_{i-1}^*|-1 
%\sum_{t=1}^{n^2} \sum_{r=1}^{n} \sum_{c=1}^{n} \sum_{\alpha=0}^3 \bar{x}_{t,r,c,%\alpha} \cdot x_{t,r,c,\alpha} \leq n^2-1
\end{equation}

\noindent
The rationale is to force at least one of the variables of set $X_{i-1}^*$ to be equal to zero.
If no solution of the previous region $R_{i-1}$ can lead to a zero lower bound in the current region $R_i$, the procedure backtracks further and searches for a new solution for region $R_{i-2}$ (and so on). 
%Because this approach is clearly much more time consuming than the pure greedy algorithm, we set the total time limit to an higher value, namely 3 hours, to compute a complete solution.
\\
\\
Due to the enumerative nature, this procedure can lead to incomplete solutions despite long computation times.
We decided to limit the backtracking procedure to a fixed time limit, after which the greedy heuristic continues until a complete solution is generated.
%Diversification is induced in a similar way as in the pure greedy heuristic.  A random inner tile is placed within the first region, before the backtracking heuristic starts.
This backtracking heuristic is sketched in Algorithm \ref{alg:backtracking}, in which a recursive method ($BACKTRACKING\_HEURISTIC$) attempts to solve the current region $R_i$, given partial solution $S^\ast_{i-1}$ 
obtained in the previous region $R_{i-1}$.
If the lower bound $lb$ of the current region  is greater than 0, the method backtracks to the previous level.
However, if the lower bound is still $0$ (and a perfectly matched assignment is found), the heuristic attempts to solve the next region. This will continue calling recursively until the puzzle is solved, or shown infeasible given the current assignments in $S^\ast_{i-1}$.
In the latter case, the current partial solution $S^\ast_{i-1}$ will be excluded and a new partial solution $S'^\ast_{i-1}$ will be constructed, different from $S^\ast_{i-1}$ and any other previously excluded partial solution.

If a timeout is reached, the method will continue with the best partial solution and solve the remaining regions with the greedy heuristic, discussed in the previous section.
%\begin{figure}[h]
%\centering
%		\includegraphics[scale=0.50]{partial6.png}
%	\caption{Partial solution obtained by the backtracking heuristic on a $6 \times 6$ puzzle using $2 \times 2$ regions.}
%\label{fig:partial6}
%\end{figure}
%
%\noindent
%Figure \ref{fig:partial6} reports an example of a partial solution constructed by the backtracking procedure.
%Only two regions of size $2 \times 2$ remain unmatched to be completed at the bottom right of the puzzle.

\begin{algorithm}
\small
\caption{$BACKTRACKING\_HEURISTIC$}
\begin{algorithmic}
\Require $\mathcal{R} = \{R_1,R_2,R_3, \ldots R_K\}$ \Comment{A decomposition of $R$ in $K$ regions $R_i$}
\Require $i$ \Comment{Current recursive level ($i=1,\ldots,K$)}
\Require $S^\ast_{i-1}: T^\ast_{i-1}\mapsto R^\ast_{i-1}$ \Comment{Partial solution of the previous level ($S^\ast_{0}$ is the initial empty partial solution)}
\State
\State $excluded solutions \gets \emptyset$ \Comment{Partial solutions not leading to feasible solutions}
\While{not timeout}
\State $S^\ast_{i} \gets OPTIMIZE\_REGION(S^\ast_{i-1},R_{i},excludedsolutions)$
\If{$lb > 0$}
\State return $S^\ast_{i-1}$ \Comment{No feasible solution in current level. (backtrack)}
\Else
\State $S^\ast_{i+1} \gets BACKTRACKING\_HEURISTIC(\mathcal{R},i+1,S^\ast_{i})$
\If{$S^\ast_{i+1} = S^\ast_{i}$} \Comment{$S^\ast_i$ does not lead to feasible solution}
\State $excluded solutions \gets excluded solutions \cup S_i^\ast$
\Else
\State $S^\ast \gets S^\ast_{i+1}$ \Comment{$S^\ast_{i+1}$ is the complete solution}
\State \Return $S^\ast$
\EndIf
\EndIf
\EndWhile
\State $S^\ast \gets GREEDY(S^\ast_i,R^\ast_i)$
\State \Return $S^\ast$
\end{algorithmic}
\label{alg:backtracking}
\end{algorithm}

\subsection{A multi-neighbourhood local search approach}
\label{subsec:VNSILS}

\noindent
A multi-neighbourhood local search approach has been developed to improve the solutions generated by the constructive heuristics (or any random solution).
The key idea is to test, after an initial, complete solution is generated by the heuristics, whether a controlled-size neighbourhood can still improve the current solution.
%If solutions generated by the two constructive heuristics are \textit{local minima} w.r.t very %large scale neighbourhoods, improvements are unlikely. 
This local search method is a \textit{Steepest Descent} search that tries to improve a solution with the following neighbourhoods: \emph{Border Optimisation, Region Optimisation, Tile Assignment} and \emph{Tiles Swap and Rotation}.
We refer to Figure \ref{fig:neighbourhoods}} for an illustration of the regions considered by these neighbourhoods.
\\
\\
The \textbf{Border Optimisation (BO)} neighbourhood only considers placing tiles in the border, while all the tiles in the inner part are fixed.
The decomposition tries to find the optimal border in terms of matching edges, also considering the fixed tiles on the adjacent inner part. Correspondingly, 
model (\ref{eq:m2obj})--(\ref{eq:varv}) is modified in such a way that 
the inner tile/positions variables are fixed to their current value.  Only the border 
tile/positions variables can change value.
This subproblem corresponds to a one-dimensional edge-matching problem.
Preliminary computational tests indicated that the related MILP model could always be solved.
Solutions for the largest instances, such as the original EII puzzle, can be generated within little computation time. 
When the \textbf{BO} neighborhood is considered, the corresponding MILP model is solved, returning a solution at least as good as the current solution and consisting of an optimal border with respect to the $(n-2)\times (n-2)$ inner region.
\\
\\
The \textbf{Region Optimisation (RO)} neighbourhood relates to the optimisation of a smaller region inside the puzzle and only considers the tiles of this region in the puzzle.
Correspondingly, given the current solution, model (\ref{eq:m2obj})--(\ref{eq:varv}) is suitably modified in such a way that the tile/position variables outside the region are fixed to their current value.
Only the region's tile/position variables can change value.
The \textbf{RO} neighborhood  can also be tackled by means of the Max-Clique formulation by generating a graph only containing nodes corresponding to tile/position assignments in the specified region. However, only feasible tile/position assignments should be considered and nodes conflicting with assignments adjacent to (but outside) the region should not be added to the graph. We recall that the purpose of the model is to find complete assignments, that is, without any unmatched edges. However, given the tiles in the considered region, it may not be feasible to find such a solution.
In this case, holes are left in the region to which the remaining, unassigned tiles should be assigned. The related MILP region model is solved where all assigned tile/position variables are fixed to the value determined by the Max-Clique solver.
When the \textbf{RO} neighborhood is considered, the local search procedure samples regions of fixed size in the current solution under consideration.
For small sizes, the Max-Clique model (heuristically) is solved faster than the MILP model.  Therefore,  
the \textbf{RO} neighborhood is always addressed by means of the Max-Clique formulation
where the MILP formulation is only used for completing the solution whenever holes are left in the region.
\\
\\
In the \textbf{Tile Assignment (TA)} neighbourhood, $k$ tiles are removed from non-adjacent positions (diagonally adjacent is allowed) and optimally reinserted, thereby minimising the number of unmatched edges.
The related subproblem corresponds to a pure bipartite weighted matching problem, which is optimally solvable by e.g.\ the Hungarian Algorithm \cite{KU55}.
The \textbf{TA} neighbourhood was first introduced by Schaus and Deville \cite{SD08} who called it a \emph{very large neighbourhood}.
Wauters et al.\ \cite{WVV12} developed a probabilistic version of the \textbf{TA} neighbourhood that sets a higher probability to selecting tiles with many unmatched edges.
The latter \textbf{TA} variant was applied in the present paper.
The \textbf{TA} separates the inner and the border moves. It is prohibited to reassign border pieces to the inner region and vice versa.
\\
\\
An extention to the \textbf{TA} neighbourhood is also tested. In particular a ``checkers'' configuration of selected tiles  is studied, i.e.~all tiles on the board that are diagonally adjacent. We denote this extension \textbf{Black and White (BW)}.
The local search procedure iterates in this neighbourhood, iteratively changing between ``black'' and ``white'' positions and solving the related bipartite weighted matching problem until no more improvements are found.
\\
\\
Finally, the \textbf{Tiles Swap and Rotation (TSR)} neighbourhood is a standard local search swap operator, in this case swapping the assignment of two tiles, trying all possible rotations as well.
The local search procedure exhaustively searches the neighbourhood until a local optimum is reached.

\begin{figure}
\centering
\subfloat[][]{\label{subfig:bo}\includegraphics[width=0.3\textwidth]{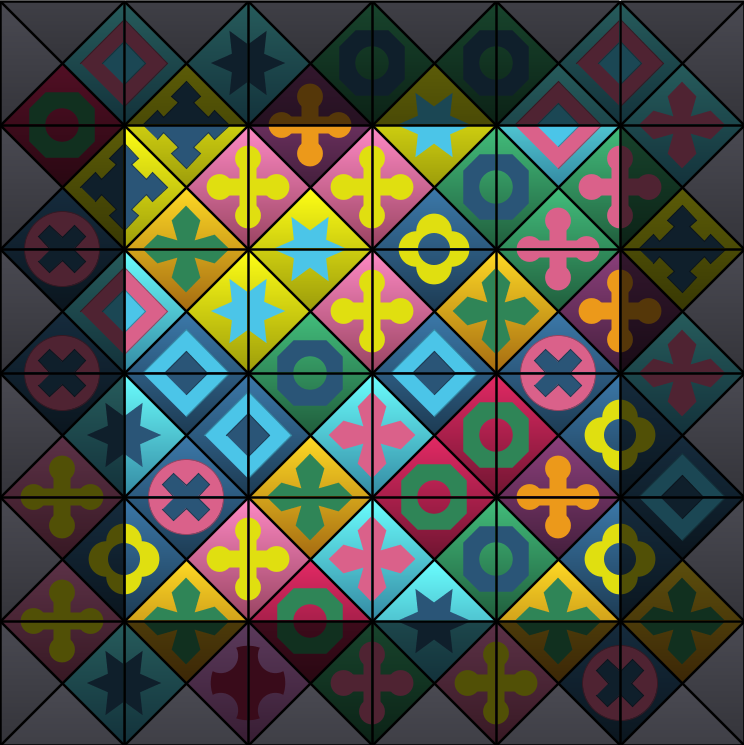}}\hfill
\subfloat[][]{\label{subfig:ro}\includegraphics[width=0.3\textwidth]{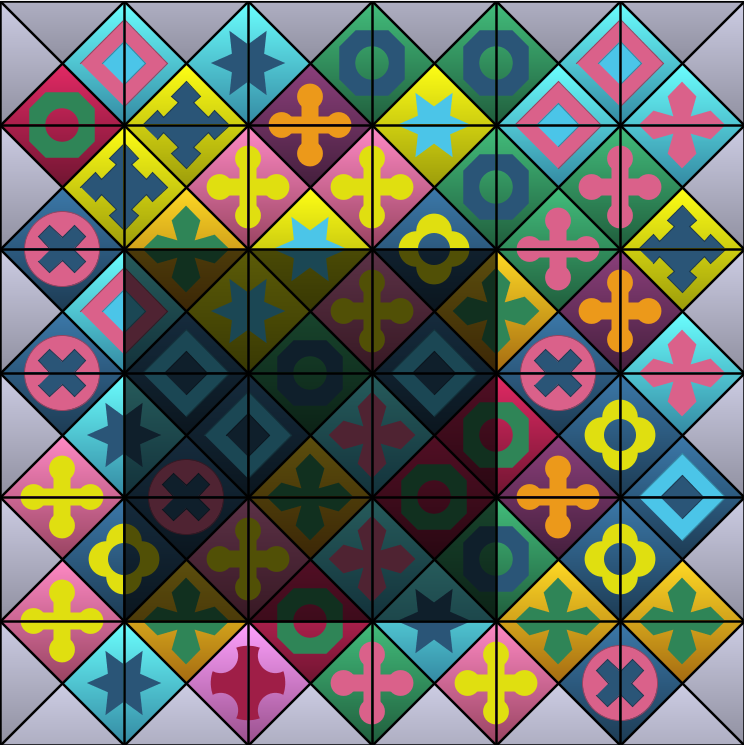}}\hfill
\subfloat[][]{\label{subfig:ta}\includegraphics[width=0.3\textwidth]{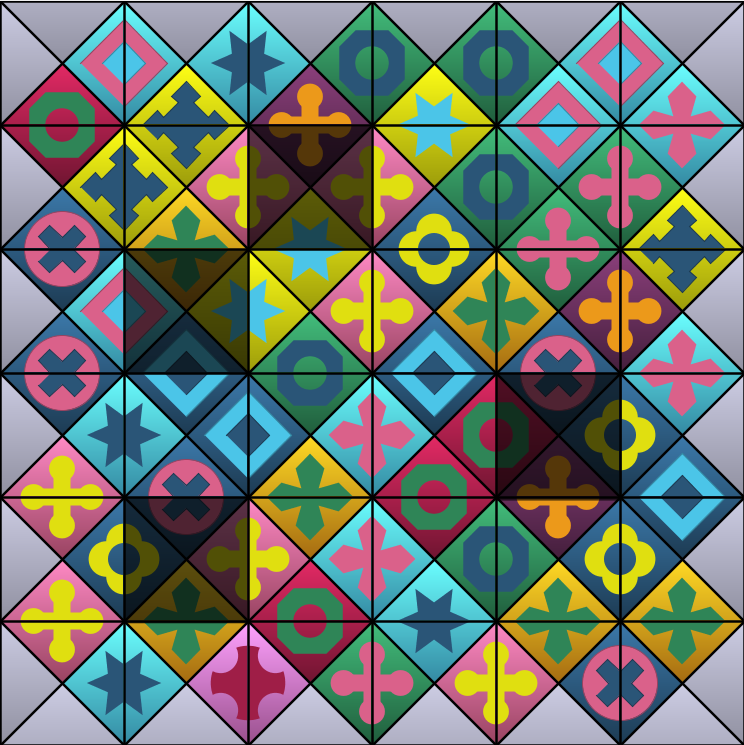}}\\
\subfloat[][]{\label{subfig:bw}\includegraphics[width=0.3\textwidth]{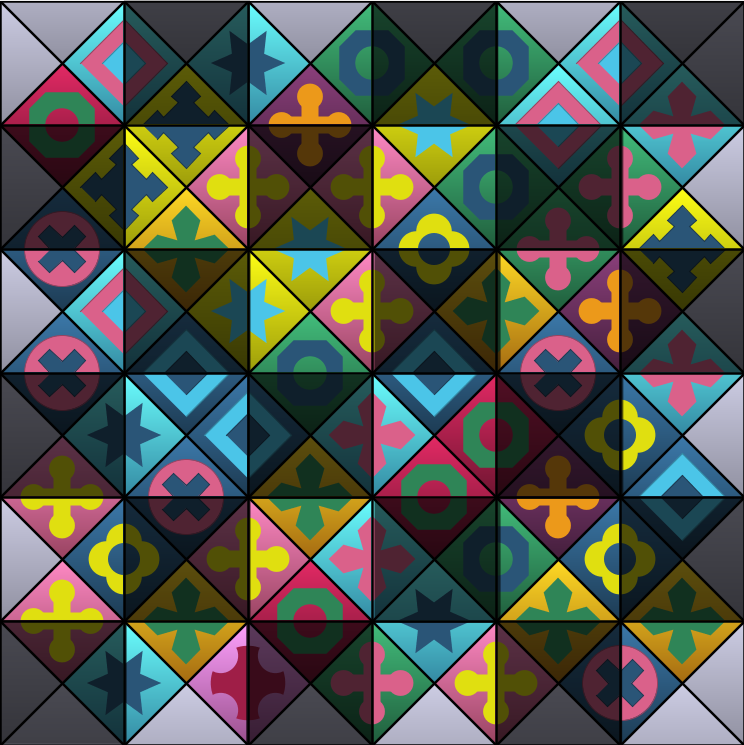}}\hspace{2em}
\subfloat[][]{\label{subfig:tsr}\includegraphics[width=0.3\textwidth]{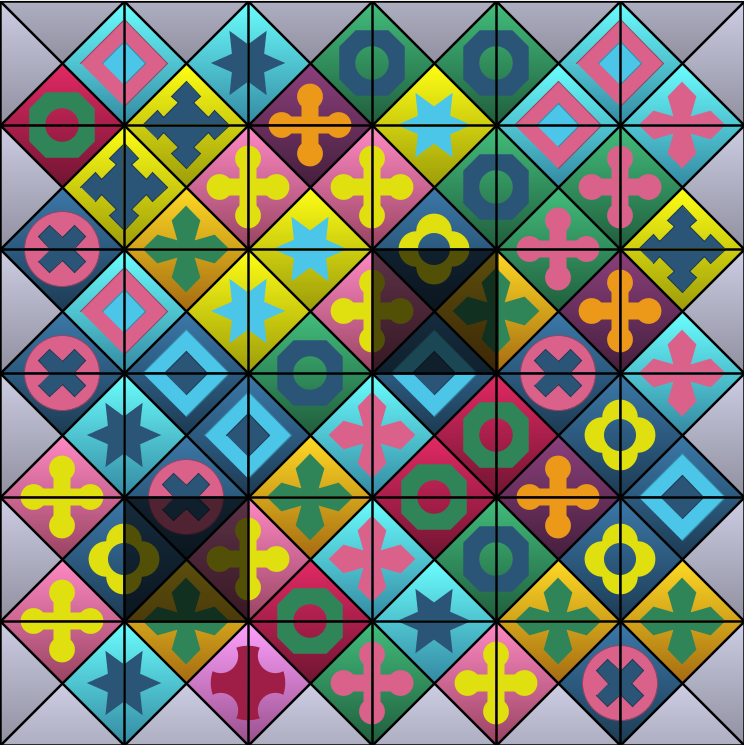}}
\caption{Illustration of the regions involved in the neighbourhood operations: \protect\subref{subfig:bo} optimizing the border (\textbf{BO}); \protect\subref{subfig:ro} optimizing a rectangular region (\textbf{RO}); \protect\subref{subfig:ta} optimizing non-adjacent tile assignments (\textbf{TA}); \protect\subref{subfig:bw} optimizing diagonally adjacent tile assignments in a \emph{checkers} fashion (\textbf{BW}); \protect\subref{subfig:tsr} swapping two tiles and possibly rotating them (\textbf{TSR}).}
\label{fig:neighbourhoods}
\end{figure}
%\subsection{A new objective function}
%\label{subsec:NewObj}
%
%Besides the neighbourhoods presented we also tested a new objective function of the local search.......

\section{Computational Results}
\label{sec:e2results}

\noindent
This section provides computational results obtained by the multi-neighbourhood local search approach on the Eternity II puzzle as well as the $10\times 10$, $12\times 12$, $14\times 14$ and $16\times 16$ instances that were used in the META 2010 EII contest\footnote{3$^{rd}$ International Conference on Metaheuristics and Nature Inspired Computing, Djerba Island, Tunisia, October 27$^{th}$-31$^{th}$ 2010 -- Eternity 2 contest \url{http://www2.lifl.fr/META10/pmwiki.php?n=Main.Contest}.}.
The latter instances serve as an interesting test set for comparison, due to the availability of some results from the contest.
In addition, to the best of our knowledge, the complete solutions of these instances are not publicly available. 
The $3\times 3$ to $9\times 9$ instances used in Section \ref{sec:models} also originate from this set.

All tests were performed on a 40 cores Intel Nehalem cluster with 120 GB ram, with each core running @ 3.46 GHz with 8 MB cache.
Computational resources provided by DAUIN's HPC Initiative\footnote{For more details see  \url{http://www.dauin-hpc.polito.it}.}.
This cluster was used to solve different instances/runs in parallel in order to reduce the total time required to run all tests. 
Each individual test was run on a single processing core, thus no parallelism was employed in the algorithms.
All MILP models are solved using CPLEX 12.4. % as MILP solver.

\bigskip

\begin{table}[h]
	\centering
		\begin{tabular}{lr}
		\hline
		\textbf{Parameter Name} & \textbf{Value}\\\hline
		$TA.K$ & 16 tiles\\
		$TA.N$ & 1000 it\\
		$Clique.W$ & 6 cols\\
		$Clique.H$ & 6 rows\\
		$Clique.N$ & 10 it\\
	%	$Shuffle.notImprovingT$& 120 s\\
%		$Shuffle.K$ & 2\\
\hline
		\end{tabular}
	\caption{Parameter Settings}
	\label{tab:param}
\end{table}
\noindent
Table \ref{tab:param} summarizes the parameter settings of the  multi-neighbourhood local search approach. The local search procedure starts either from a random solution or from a solution obtained by the constructive heuristics. The algorithm cycles through the proposed neighbourhoods in the following sequence: \textbf{TA} (for $TA.N$ iterations with sample size $TA.K$), \textbf{BO} (for one iteration), \textbf{BW} (till local optimum), \textbf{TSR} (till local optimum) and finally \textbf{RO} by means of the Max-Clique formulation (for $Clique.N$ iterations with rectangular sample size $Clique.W\times Clique.H$). 
This sequence was determined experimentally, though the difference in performance between sequences was very limited.
At the end of the multi-neighbourhood step, the final solution is a local minimum with respect to 
all considered neighbourhoods.
%The procedure keeps track of the elapsed time since the last improvement. If this exceeds $Shuffle.notImprovingT$ then the algorithm shuffles the current solution with $Shuffle.K$ random swaps.
\\
\\
Table \ref{tab:greedyconstruct} shows the results for twenty runs of the MILP-based greedy heuristic and the MILP-based backtracking heuristic for different region sizes on all the problem instances.
A timeout of $1200$ seconds was set for the backtracking heuristic for the $10\times 10$ instance, $1800$ seconds for the $12\times 12$ instance, $2400$ seconds for the $14\times 14$ instance and $3600$ seconds for the $16\times 16$ and EII instances.
The greedy heuristic (executed by itself, or after the backtracking heuristic) is executed until all regions are solved.
The table also shows the results of both constructive methods after subsequent optimisation by the local search heuristic.\\
In general, both constructive heuristics generate better results when larger regions are used. 
This clearly affects the CPU time needed to compute optimal solutions for each region. %, as can be drawn from the entries related to the CPU time column of the table.
%It seems that more than 24 tiles lead to a definitely strong increased time for solving the %complete puzzle.
By comparing the results of the two heuristics (without the local search phase), it seems that the backtracking procedure does not strongly dominate (on maximum, average and minimum values) the greedy one, while consuming all the available time.
This dominance tends to be more evident for small puzzles and region sizes, while for larger instances with solutions generated by larger regions, the gap becomes smaller.
\\
\\
In almost all cases, the local search procedure manages to improve the results of the constructive heuristics by several units, %This usually occurs before shuffling
indicating that these initial solutions are not local optima with respect to the considered neighbourhoods. We conclude that  many neighbourhoods in a complex structure are effective for improving these greedy constructive solutions.
Table \ref{tab:localsearchonly} shows the performance of the local search procedure starting from a random initial solution of poor quality. 
The procedure can achieve good quality results for the $10\times 10$ instance, but not for larger instances. This can easily be related to the size of the \textbf{RO} neighborhood.
As it is quite large with respect to the puzzle size in the $10\times 10$ instance, it is able to optimize a large part of the puzzle. However, this ratio becomes smaller and is thus less effective for larger instances.
\\
\\
Finally, Table \ref{tab:comparison} compares the best published results with the results obtained by the hybrid local search procedure. 
The CPU times refer to the considered time limits. 
Table \ref{tab:comparison} also reports a large test of the procedure, where the best performing configuration (backtracking+LS) was run 100 times within a doubled execution time limit. Larger execution times (using CPLEX 12.4 as ILP solver)
do not induce further improvements of the results. 
We note that some of the entries of Table \ref{tab:comparison} are missing. Many of the approaches only deal with a subset of the considered instances.
Only three studies \cite{CCCHSO10,WC10,WVV12} report results for the $10\times 10$ to $16\times 16$ instances that were tested in this paper.
Some approaches \cite{SD08,MGS09,VE08} were only applied to the real EII game puzzle.

The algorithm reported in \cite{SD08} was executed on a CPU Intel Xeon(TM)
2.80GHz, with computation time 24 hours. The best score over 20 runs equals 458.\\
\cite{MGS09} obtained a best score of 371. No indication was provided on the computer and the required CPU time.\\
The algorithm of \cite{CCCHSO10} 
was run on a PC Pentium Core 2 Quad (Q6600), 2.4 GHz, with 8 GB of RAM.  It
 considered EII style problems (but not the real puzzle) with sizes 
$10\times 10$, $12\times 12$, $14\times 14$ and $16\times 16$.
The corresponding time limits were 1200, 1800, 2400 and 3600 seconds respectively
and the entries of Table \ref{tab:comparison} report the best solution obtained over 30 runs.\\
The algorithm of \cite{WC10} addressed the same instances with the same time limits and number of runs as \cite{CCCHSO10}. It was tested on a 
personal computer with 1.8GHz CPU and 1GB RAM.
Time limits and number of runs were the same as in \cite{WC10}. The tests were performed on an  Intel Core 2 Duo @ 3Ghz with 4GB of RAM.\\
The algorithm of \cite{WVV12} was tested on all the instances from 
\cite{CCCHSO10} and \cite{WC10} and also on the  EII real game puzzle.
The entry reports the best result obtained over 30 runs with a time limit of 3600 seconds for EII.\\
Finally, the algorithm of \cite{VE08} was tested on the  EII real game puzzle only running on a grid computing system
over a period of several weeks/months not explicitly indicated by the authors.
\\
\\
The results show that the algorithm is competitive with the state of the art, obtaining top results for the $10\times 10$  instance in a similar time frame as the other algorithms.
Most interesting, the initial solution constructed by the greedy and backtracking heuristics is already of high quality, leaving only a limited gap from the optimal solution.
Therefore, we expect that these methods may serve as the basis for reaching new top results.
The best result for the official EII puzzle instance, 467 obtained using a slipping tile, scan-row backtracking algorithm \citep{VE08}, is still out of our current grasp.
However, that algorithm was highly tailored to the EII puzzle instances, used precomputed sequences and was run over the course of several weeks/months 
(see 
http:$//$www.shortestpath.se$/$eii$/$eii$\_$details.html).
A direct comparison with the approach presented is partially misleading.
Among the other existing approaches, only \cite{WVV12} shows to be slightly superior to our approach.
However, our approach should become more competitive along with the expected performance improvement of MILP solvers over the years.
Clearly, solving larger subregions in both the constructive heuristics (greedy and backtracking) will lead to better initial solutions. 
In addition, the effectiveness of the MIP-based local search neighbourhoods is expected to improve when larger regions can be solved.
If performance improvements allow ILP solvers to address instances of size $8 \times 8$ or even $9 \times 9$ in a reasonable amount of time, it may safely be assumed that the proposed approach will lead to improved results competitive with the other state of the art approaches.

%\begin{landscape}
\begin{table}[h]
\scriptsize
	\centering
		\begin{tabular}{| l l | c c c c c |}
		\hline %\hline												
\textbf{INSTANCE}	 & 	\textbf{START}	 & 	\textbf{Region Size}	 & 	\textbf{MAX}	 & 	\textbf{AVG}	 & 	\textbf{MIN}	 & 	\textbf{Time Avg. (s)}	\\\hline \hline
\textbf{10x10}	 & 	greedy	 & 	1 x 10	 & 	165	 & 	161.10	 & 	158	 & 	4.29	\\
	 & 	greedy+LS	 & 	1 x 10	 & 	168	 & 	164.90	 & 	161	 & 	1204.29	\\
	 & 	greedy	 & 	2 x 10	 & 	170	 & 	166.35	 & 	164	 & 	25.93	\\
	 & 	greedy+LS	 & 	2 x 10	 & 	170	 & 	167.05	 & 	164	 & 	1225.93	\\
	 & 	backtracking	 & 	1 x 10	 & 	169	 & 	165.65	 & 	162	 & 	1200.24	\\
	 & 	backtracking+LS	 & 	1 x 10	 & 	172	 & 	167.75	 & 	165	 & 	2400.24	\\
	 & 	backtracking	 & 	2 x 10	 & 	170	 & 	167.65	 & 	164	 & 	1207.06	\\
	 & 	backtracking+LS	 & 	2 x 10	 & 	171	 & 	168.10	 & 	165	 & 	2407.06	\\\hline
\textbf{12x12}	 & 	greedy	 & 	1 x 12	 & 	245	 & 	241.20	 & 	239	 & 	6.83	\\
	 & 	greedy+LS	 & 	1 x 12	 & 	247	 & 	244.00	 & 	240	 & 	1806.83	\\
	 & 	greedy	 & 	2 x 12	 & 	249	 & 	247.00	 & 	244	 & 	101.02	\\
	 & 	greedy+LS	 & 	2 x 12	 & 	250	 & 	247.50	 & 	245	 & 	1901.02	\\
	 & 	backtracking	 & 	1 x 12	 & 	248	 & 	245.35	 & 	242	 & 	1801.51	\\
	 & 	backtracking+LS	 & 	1 x 12	 & 	250	 & 	247.75	 & 	246	 & 	3601.51	\\
	 & 	backtracking	 & 	2 x 12	 & 	249	 & 	247.60	 & 	246	 & 	1868.85	\\
	 & 	backtracking+LS	 & 	2 x 12	 & 	252	 & 	248.45	 & 	246	 & 	3668.85	\\\hline
\textbf{14x14	} & 	greedy	 & 	1 x 14	 & 	338	 & 	333.00	 & 	329	 & 	10.43	\\
	 & 	greedy+LS	 & 	1 x 14	 & 	339	 & 	335.56	 & 	332	 & 	2410.43	\\
	 & 	greedy	 & 	2 x 14	 & 	344	 & 	340.50	 & 	338	 & 	2335.42	\\
	 & 	greedy+LS	 & 	2 x 14	 & 	344	 & 	340.80	 & 	338	 & 	4735.42	\\
	 & 	backtracking	 & 	1 x 14	 & 	342	 & 	338.25	 & 	334	 & 	2401.36	\\
	 & 	backtracking+LS	 & 	1 x 14	 & 	344	 & 	340.69	 & 	336	 & 	4801.36	\\
	 & 	backtracking	 & 	2 x 14	 & 	344	 & 	342.17	 & 	340	 & 	4664.65	\\
	 & 	backtracking+LS	 & 	2 x 14	 & 	344	 & 	342.33	 & 	340	 & 	7064.65	\\\hline
\textbf{16x16	} & 	greedy	 & 	1 x 16	 & 	448	 & 	444.05	 & 	440	 & 	24.08	\\
	 & 	greedy+LS	 & 	1 x 16	 & 	449	 & 	446.20	 & 	443	 & 	3624.08	\\
	 & 	greedy	 & 	2 x 16	 & 	454	 & 	451.38	 & 	448	 & 	11188.03	\\
	 & 	greedy+LS	 & 	2 x 16	 & 	454	 & 	451.88	 & 	448	 & 	14788.03	\\
	 & 	backtracking	 & 	1 x 16	 & 	453	 & 	448.80	 & 	444	 & 	3605.68	\\
	 & 	backtracking+LS	 & 	1 x 16	 & 	454	 & 	451.53	 & 	449	 & 	7205.68	\\
	 & 	backtracking	 & 	2 x 16	 & 	457	 & 	453.69	 & 	451	 & 	13722.86	\\
	 & 	backtracking+LS	 & 	2 x 16	 & 	457	 & 	454.00	 & 	451	 & 	17322.86	\\\hline
\textbf{EII-instance}	 & 	greedy	 & 	1 x 16	 & 	449	 & 	443.75	 & 	440	 & 	21.85	\\
	 & 	greedy+LS	 & 	1 x 16	 & 	450	 & 	446.50	 & 	442	 & 	3621.85	\\
	 & 	greedy	 & 	2 x 16	 & 	456	 & 	451.75	 & 	450	 & 	13835.04	\\
	 & 	greedy+LS	 & 	2 x 16	 & 	457	 & 	451.95	 & 	450	 & 	17435.04	\\
	 & 	backtracking	 & 	1 x 16	 & 	453	 & 	449.00	 & 	446	 & 	3605.08	\\
	 & 	backtracking+LS	 & 	1 x 16	 & 	454	 & 	451.35	 & 	447	 & 	7205.08	\\
	 & 	backtracking	 & 	2 x 16	 & 	457	 & 	452.80	 & 	448	 & 	15294.52	\\
	 & 	backtracking+LS	 & 	2 x 16	 & 	457	 & 	453.15	 & 	449	 & 	18894.52	\\\hline
%	\hline
\end{tabular}
	\caption{Results for the MILP-based greedy and backtracking heuristics with and without local search, using different region sizes.}
	\label{tab:greedyconstruct}

			\begin{tabular}{l c c c c c}
		\hline
		&	$\mathbf{10 \times 10}$ & $\mathbf{12 \times 12}$ & $\mathbf{14 \times 14}$ & $\mathbf{16 \times 16}$ & $\mathbf{EII-instance}$\\
		\hline
		$\mathbf{Obj_{Max}}$ & 167 & 238 & 312 & 398 & 391 \\
$\mathbf{Obj_{Avg}}$ & 163.2 & 231.1 & 299.2 & 384.3 & 382 \\
$\mathbf{Obj_{Min}}$ & 158 & 223 & 292 & 367 & 372 \\
	\hline
		\textbf{Optimum} & 180 & 264 & 364 & 480 & 480\\
	\hline
		\end{tabular}
	\caption{Results for the local search procedure with optimal neighbourhoods starting from a random solution.}
	\label{tab:localsearchonly}
	\end{table}

	\begin{table}
	\scriptsize
		\centering
		\begin{tabular}{l c c c c c}
		\hline
		&	$\mathbf{10 \times 10}$ & $\mathbf{12 \times 12}$ & $\mathbf{14 \times 14}$ & $\mathbf{16 \times 16}$ & $\mathbf{EII-instance}$\\
		\hline
%	\textbf{Present paper (20 runs)}	 & 172 (20 min)  & 252 (30 min 26 s.) & 344 (9min 28 s.) & 457 (2 h. 56 min) & 457 (2 h. 36 min)\\
%	\textbf{Present paper (100 runs)}	 & 172   & 252  & 347 & 457  & 458 \\\hline
	\textbf{Present paper (20 runs)}	 & 172 (20 min)  & 252 (30 min) & 344 (40 min) & 457 (180 min) & 457 (180 min)\\
	\textbf{Present paper (100 runs)}	 & 172 (40 min)  & 252 (60 min)& 347 (80 min) & 457 (360 min)  & 458 (360 min) \\\hline
	Mu\~noz et al.\ \cite{MGS09} 							 & - & - & - & - & 371 (-) \\
	Wang and Chiang \cite{WC10} & 163 (20 min)  & 234 (30 min) & 318 (40 min) & 418 (60 min) & -\\
	Coelho et al.\ \cite{CCCHSO10} & 167 (20 min)  & 241 (30 min) & 325 (40 min) & 425 (60 min) & -\\
	Schaus and Deville \cite{SD08} &  - &  - & - & - & 458 (1440 min)\\
	Wauters et al.\ \cite{WVV12} & 172 (20 min)  & 254 (30 min) & 348 (40 min) & 460 (60 min) & 461 (60 min)\\
	Verhaard \cite{VE08} & - & - & - & - & 467 (weeks/months)\\
	\hline
	\textbf{Optimum} & 180 & 264 & 364 & 480 & 480\\
	\hline
		\end{tabular}
	\caption{Comparison of the best results to other approaches available in the literature. (Execution times presented within parenthesis)}
	\label{tab:comparison}
	\end{table}

%\end{landscape}

\section{Conclusions}
\label{sec:e2conclusion}

\noindent
The present work introduced a hybrid approach to the Eternity II puzzle. 
A MILP formulation is related to the puzzle's optimisation version, where the total number of unmatched edges should be minimised.
It is shown that the Eternity II puzzle can be modelled as a \textit{clique} problem, providing, as a byproduct of this work, new hard instances of the maximum clique problem to the community.
Preliminary testing revealed it was clear that these models cannot handle large size instances (such as the original EII puzzle), as they quickly become computationally intractable.
Therefore, these models were used as the basis for heuristic decompositions, which could then be used in a hybrid approach.
\\
\\
A greedy and a backtracking constructive heuristic have been designed, which strongly rely on the capability of optimally solving a specific region of the puzzle.
Within a reasonable time limit, high quality solutions can be generated using these heuristics.
A multi-neighbourhood local search approach has also been proposed.
By applying a set of different neighbourhoods, 
%of which two are based on decompositions from the MILP and clique model, and two are very %large scale neighbourhoods, 
the local search procedure manages to improve upon the initial solutions generated by the constructive heuristics and reaches solutions competitive to the best available results. 
\\
\\
These results confirm that a novel and clever use of mathematical models and solvers/heuristics is effective for large size problems, which cannot be solved all in once by the same MILP solver.
We believe that 
hybridizing local search approaches and mathematical programming techniques in a matheuristic context
is the key to break up the intractability of hard problems such as the EII puzzle.

%\bibliographystyle{model3-num-names}
%\bibliography{e2-bibfile}

%% not want to use model3-num-names.bst.


\end{document}